\documentstyle[12pt,aaspp4,flushrt,amssym]{article}

\begin{document}

\slugcomment{To appear in The Astrophysical Journal}

\title{The Luminosity Function of Field Galaxies 
         in the CNOC1 Redshift Survey} 
\author{H. Lin \altaffilmark{1}, 
        H. K. C. Yee \altaffilmark{1,2}, 
        R. G. Carlberg \altaffilmark{1,2}, and
        E. Ellingson \altaffilmark{2,3}}

\altaffiltext{1}{Department of Astronomy, University of
    Toronto, 60 St. George St., Toronto, ON M5S 3H8, Canada, 
    lin@astro.utoronto.ca, 
    hyee@astro.utoronto.ca, carlberg@astro.utoronto.ca}

\altaffiltext{2}{Visiting Astronomer, Canada-France-Hawaii
    Telescope, which is operated by the National Research Council of
    Canada, the Centre Nationale de la Recherche Scientifique of
    France, and the University of Hawaii.}

\altaffiltext{3}{Center for Astrophysics and Space Astronomy,
  University of Colorado, CO 80309, USA, e.elling@casa.colorado.edu}

\begin{abstract}

We have computed the luminosity function for a sample of 389
field galaxies from the Canadian Network for Observational Cosmology
cluster redshift survey (CNOC1), over the redshift
range $z = 0.2-0.6$. We find Schechter parameters 
$M^*_r - 5 \log h = -20.8 \pm 0.4$ and 
$\alpha = -1.3 \pm 0.2$ in rest-frame Gunn $r$, and
$M^*_{B_{AB}} - 5 \log h = -19.6 \pm 0.3$ and 
$\alpha = -0.9 \pm 0.2$ in rest-frame $B_{AB}$.
We have also split our sample at the color of a redshifted but nonevolving Sbc galaxy, 
and find distinctly different luminosity functions for red and blue galaxies. 
Red galaxies have a shallow slope
$\alpha \approx -0.4$ and dominate the bright end of the luminosity
function, while blue galaxies have a steep $\alpha \approx -1.4$ 
and prevail at the faint end. 
Comparisons of the CNOC1 results to analogous intermediate-redshift 
luminosity functions from the Canada-France (CFRS) and Autofib redshift surveys 
show broad agreement among these independent samples, but there are also
significant differences which will require larger samples to resolve.
Also, in CNOC1 the red galaxy
luminosity density stays about the same over the range $z = 0.2-0.6$,
while the blue galaxy luminosity density increases steadily with redshift.
These results are consistent with the trend of the luminosity
density vs. redshift relations seen in the
CFRS, though the normalizations of the luminosity densities 
appear to differ for blue galaxies.
Comparison to the {\em local} luminosity function from
the Las Campanas redshift survey (LCRS) shows that the luminosity density at
$z \approx 0.1$ is only about half that seen at $z \approx 0.4$. 
A change in the luminosity function shape, particularly at the faint end, 
appears to be required to match the CNOC1 and LCRS luminosity
functions, if galaxy evolution is the sole cause of the differences seen.
However, it should be noted that the specific details of the
construction of different surveys may complicate the comparison of
results and so may need to be considered carefully.

\end{abstract}


\section{Introduction}

The luminosity function of galaxies is a simple but fundamentally
important quantity in the study of galaxy populations and their
evolution. The luminosity function is of particular relevance to the
problem of the excess counts of faint blue galaxies (e.g., \cite{koo92};
\cite{koo96}) and will provide important constraints toward the
resolution of this question. In particular, accurate determinations of
the luminosity function at both low and high redshifts are crucial.
Large wide-angle redshift surveys are providing precise measurements
of the luminosity function in the local $z \sim 0$ universe (e.g.
\cite{lin96lum}; \cite{marz94}; \cite{lov92}), a necessary baseline on
which to anchor models of galaxy evolution back to higher redshifts.
Also, recent smaller but deeper redshift surveys have provided direct
measurements of the luminosity function to redshifts $z \approx 1$
(e.g. \cite{lil95}; \cite{ell96}; \cite{cow96}; \cite{gla95}), and
have revealed clear evidence for the evolution of the luminosity
function with lookback time. Moreover, that evolution depends strongly
on galaxy type, such that the luminosity density of blue, star-forming
galaxies appears to have increased substantially by $z \sim 0.5$,
whereas that of red, more quiescent galaxies appears to have changed
relatively little. 

In this paper we present the luminosity function for a sample of field
galaxies obtained as part of the 
Canadian Network for Observational Cosmology
cluster redshift survey (CNOC1). Although the CNOC1 survey was optimized
for obtaining cluster galaxy redshifts, a concurrent sample of field
galaxies was an important and necessary component needed in order to
accomplish the main survey goal of an accurate
measurment of $\Omega$ from cluster dynamics (\cite{car96}). The 389
galaxies in the CNOC1 field sample considered in this paper span the
redshift range $z = 0.2 - 0.6$, and our sample size is comparable to
that of other surveys at these intermediate redshifts. Moreover,
available color information allows us to subdivide our sample by
galaxy type. The aim of this paper is to compute the luminosity
function for the CNOC1 field galaxies and to compare our results to
those from other intermediate-redshift surveys. 
We describe the CNOC1 data sample in \S~\ref{data} and detail our methods 
in \S~\ref{methods}. Our luminosity function results are then 
presented and discussed in \S~\ref{results}.
We summarize our conclusions in \S~\ref{conclusions}.

\section{The CNOC1 Survey Data} \label{data}

The CNOC1 cluster redshift survey contains about 2600 velocities
of cluster and field galaxies, observed in the fields of 16 high X-ray
luminosity clusters spanning the redshift range $z \approx 0.2 - 0.6$.
Photometric and spectroscopic observations were obtained using the
Multi-Object Spectrograph (MOS) at the Canada-France-Hawaii Telescope
(CFHT) during 24 nights in 1993 and 1994. A detailed description of
the observational and data reduction techniques is given in Yee,
Ellingson, \& Carlberg (1996; hereafter YEC); here we briefly describe
some relevant details. 

The CNOC1 field sample used in this paper is defined to lie within the
redshift limits $0.2 < z < 0.6$, and outside the individual CNOC1 cluster
redshift limits given in Table~1 of Carlberg et al.\ (1996).
In addition, further redshift limits are introduced by the use of four
band-limiting filters, designed to optimize galaxy spectroscopy for
clusters in different redshift ranges (YEC).
For each filter, we set the low redshift limit where [OII] $\lambda
3727$ would be 50 \AA \ from the filter's blue end, and we set the
high redshift limit where G-band would be 150 \AA \ from the filter's
red end. 
(Note our high redshift filter limits are more conservative
than the corresponding limits given in YEC, which were based on
detectability of Ca H+K. The addition of G-band here and the resulting
more stringent limits should minimize the effect of any possible 
incompleteness at the high-$z$ ends of the filters, which we are still
investigating.)
Selection of objects for spectroscopy was based on Gunn $r$
magnitudes (\cite{thu76}). In order to optimize the number of cluster redshifts, 
the fraction of galaxies spectroscopically observed was designed to decline with
apparent magnitude. Thus a magnitude
selection weight $w_m$ (where $1 / w_m$ is the fraction of objects with
redshifts at a given apparent magnitude; see YEC) is assigned to each
galaxy to properly weight 
its contribution in any statistical analyses (see below). In this paper we
limit the sample to apparent magnitudes for which $w_m \lesssim 5$.
Gunn $g$ magnitudes are also derived for each
galaxy, and we further use a color selection weight $w_c (g-r)$
(\cite{elling96}) to remove any residual
color-dependent spectroscopic sampling effects not accounted
for by the main magnitude selection weight. However, our results are
little changed by inclusion of $w_c$. 
We use the observed $g-r$ colors and interpolation among
model galaxy spectral energy distributions (\cite{cww})
to derive rest-frame colors and appropriate $k$-corrections for determination
of absolute magnitudes $M$, in both the Gunn $r$ and $B_{AB}$
(\cite{oke72}) systems. (For this purpose, we also need the
transformations $r_{AB} = r - 0.21$ and $g_{AB} = g + 0.05$ given by
Fukugita et al.\ 1995.) Throughout this paper we adopt a deceleration
parameter $q_0 = 0.5$ and a Hubble constant $H_0 = 100 \
h$~km~s$^{-1}$~Mpc$^{-1}$; we use $h = 1$ if not otherwise indicated.
Finally, we limit the absolute magnitude range to $-22.0 <
M_r < -17.5$ or $-21.5 < M_{B_{AB}} < -17.0$, 
outside of which we have few observed galaxies. 
The final field sample consists of 389 galaxies.
Table~\ref{tabsamps} summarizes the numbers, redshift limits, and
apparent magnitude limits of our data sample.

\section{Methods} \label{methods}

We compute the luminosity function using two related methods which are
unbiased by density inhomogeneities in the galaxy distribution. These are
the parametric 
maximum-likelihood method of Sandage, Tammann, \& Yahil (1979; 
hereafter STY) and the non-parametric stepwise maximum-likelihood
(SWML) method of Efstathiou, Ellis, \& Peterson (1988; hereafter
EEP). We briefly describe these methods below. Fuller descriptions
may be found in the STY and EEP papers, as well as in Lin et al.\ (1996a).

Consider a galaxy $i$ observed at redshift $z_i$ in a flux-limited redshift
survey.  Let $m_{{\rm min},i}$ and $m_{{\rm max},i}$ denote 
the apparent
magnitude limits of the field in which galaxy $i$ is located, and also
impose absolute magnitude limits $M_1 < M < M_2$ on the sample. Let $\phi(M)$
be the differential galaxy luminosity function which we want to determine. 
Then the probability that 
galaxy $i$ has absolute magnitude $M_i$ is given by
\begin{equation} \label{eqprob}
p_i \equiv p(M_i \vert z_i) \propto
    \phi(M_i) \left/ \int^{\min[M_{\rm max}(z_i), M_2]}_{\max[M_{\rm min}(z_i), M_1]}
                     \phi(M) dM \right. \ ,
\end{equation}
where $M_{\rm min}(z_i)$ and $M_{\rm max}(z_i)$ denote the
absolute magnitude limits, at $z_i$, corresponding to the 
given apparent magnitude limits.
We can then form a likelihood function, $\frak L$, for having a survey
of $N$ galaxies,
with respective absolute magnitudes $M_i$, by multiplying the 
individual probabilities $p_i$ together, obtaining
\begin{equation} \label{eqlnlike}
\ln \frak L = \sum_{i=1}^{N} \left\{ 
                         \ln \phi(M_i) - 
                         \ln \int^{\min[M_{\rm max}(z_i), M_2]}_{\max[M_{\rm min}(z_i), M_1]}
                           \phi(M) dM
                       \right\} W_i + {\rm constant} \ .
\end{equation}
Here the weight $W_i$ is a modification needed for the CNOC1 survey
in order to account for the apparent magnitude and color
selection effects described in \S~\ref{data}: $W_i = w_{m,i} w_{c,i}$
(cf. \cite{zuc94}).

In the STY method one assumes a parametric model for $\phi(M)$, and the
parameters describing $\phi(M)$ are determined by maximizing the likelihood 
$\frak L$, or equivalently $\ln \frak L$,
with respect to those parameters. In our case we take as our model for
$\phi(M)$ the Schechter function (\cite{sch76})
\begin{equation} \label{eqschech}
\phi(M) = (0.4 \ln 10) \ \phi^* \ [10^{0.4(M^*-M)}]^{1+\alpha} \
                         \exp [-10^{0.4(M^*-M)}] \ ,
\end{equation}
and use the STY method to find the characteristic magnitude $M^*$ and the 
faint-end slope $\alpha$. We emphasize that in the STY method, the
determination of $M^*$ and $\alpha$ are unbiased by the presence of
galaxy density fluctuations in the survey, unlike the case for
traditional least-squares luminosity function estimators (e.g., \cite{fel77}).
On the other hand, the normalization $\phi^*$ drops out in 
equation~(\ref{eqprob}) and has to be determined separately as 
described below. 
Error ellipses in the $M^*$-$\alpha$ plane 
may be drawn by finding the contour corresponding to
\begin{equation} \label{ellipse}
\ln \frak L = \ln \frak L_{\rm max} - \case{1}{2} \Delta \chi^2 \ ,
\end{equation}
where $\Delta \chi^2$ is the change in $\chi^2$ appropriate for the 
desired confidence level
and for a $\chi^2$ distribution with 2 degrees of freedom 
(e.g., $\Delta \chi^2 = 6.17$ at $2\sigma$ confidence).

Alternatively, one does not have to assume a particular functional form
for $\phi(M)$. Rather, in the EEP stepwise maximum likelihood method (SWML), 
the luminosity function is taken to be a series of $N_p$ steps, each of width
$\Delta M$ in absolute magnitude:
\begin{equation}
\phi(M) = \phi_k \ , \ \ M_k - \Delta M / 2 < M < M_k + \Delta M / 2 \ , \ \
                     k = 1, \ \ldots , \ N_p \ .
\end{equation}
Here the likelihood is maximized with respect to the 
$\phi_k$, which are solved for via an iterative procedure.
Also, we can estimate the variances of the $\phi_k$ by calculating
the diagonal elements of an appropriate covariance matrix. 
See EEP for details of the procedures used to estimate the $\phi_k$
and their errors.

To calculate the normalization $\phi^*$ of the Schechter function
(\ref{eqschech}), as well as the mean galaxy number density $\bar\rho$, we do
the following.
For galaxies $i$ within redshift
limits $z_1 < z_i < z_2$ and absolute magnitude limits $M_1 < M_i <M_2$, we
estimate $\bar\rho$ by 
\begin{equation} \label{eqrho}
\bar\rho = \frac{ \sum_i W_i / S(z_i) }
                { V } \ ,
\end{equation} 
where $V$ is the appropriate (comoving) volume, 
and $S(z)$ is the selection function defined by
\begin{equation} \label{selfunc}
S(z) = \int^{\min[M_{\rm max}(z), M_2]}_{\max[M_{\rm min}(z), M_1]}
            \phi(M) dM
     \left/ 
       \int^{M_2}_{M_1} \phi(M) dM
     \right. \ .
\end{equation}
The Schechter function normalization $\phi^*$ is then just
\begin{equation}
\phi^* = \frac{ \bar\rho }
              { \int^{M_2}_{M_1} \phi'(M) dM } \ ,
\end{equation}
where $\phi'$ is the Schechter function with $\phi^*$ set to one. 
We use bootstrap resampling (e.g., \cite{bar84}) to estimate the errors 
on $\bar\rho$ and $\phi^*$ contributed by uncertainties from fitting $M^*$ and 
$\alpha$, and from sampling and weighting fluctuations. This method does
not include errors arising from galaxy density fluctuations, 
which we instead estimate using an appropriate integral over the galaxy
clustering power spectrum $P(k)$:
\begin{equation}
(\delta \bar\rho / \bar\rho)^2 = \frac{1}{(2\pi)^3}
   \int d^3k \ P(k) \ |W({\bf k})|^2.
\end{equation}
Here $V$ denotes the volume of the particular sample under  
consideration, and $W({\bf k})$ is the window function for that
volume, defined by
\begin{equation}
W({\bf k}) \equiv \frac{1}{V} \int_V d^3{\bf r}\ e^{i{\bf k} \cdot {\bf r}}\ .
\end{equation}
(That is, in $V$ we explicitly account for the ``pencil-beam''
geometry of the CNOC1 fields.)
For $P(k)$ we adopt the local result derived from a 19000-galaxy sample drawn
from the Las Campanas Redshift Survey (LCRS; \cite{lin96ps}), but
multiplied by $1 / (1+z)^2$ to account for linear evolution
(e.g., \cite{pee93}, pp. 528-529)
at the higher redshifts sampled in CNOC1. The final errors on $\phi^*$
and $\bar\rho$ consist of the quadrature sum of the bootstrap
resampling and density fluctuation errors; it turns out for our sample
that these two sources of error make roughly equal contributions.
We will also calculate luminosity densities $\rho_L$ for our
galaxies, using
\begin{equation}
\rho_L = \frac{\sum_i W_i \ 10^{-0.4 M_i} / S_L(z_i)}{V} \ ,
\end{equation}
where
\begin{equation} \label{eqsl}
S_L(z) = \int^{\min[M_{\rm max}(z), M_2]}_{\max[M_{\rm min}(z), M_1]}
            10^{-0.4 M} \phi(M) dM
     \left/ 
       \int^{M_{\rm faint}}_{-\infty} 10^{-0.4 M} \phi(M) dM
     \right. \ ;
\end{equation} 
that is, we sum over the luminosities of our observed galaxies, but
weighted by the factor $S_L(z)$, which uses the luminosity function 
$\phi$ to extrapolate
for the luminosity of unobserved galaxies lying outside the accessible
survey flux limits. Note we use a finite $M_{\rm faint}$ 
($-17 + 5\log h$ for both $r$ and $B_{AB}$ bands) instead
of $+\infty$ to prevent the denominator of $S_L(z)$ from blowing up
when $\alpha \leq -2$, as sometimes occurs during bootstrap
resamplings of our blue galaxy subsample (\S~\ref{results}). Also,
note that our survey apparent flux limits prevent us from measuring
the luminosity function fainter than our chosen $M_{\rm faint}$
(c.f. end of \S~\ref{data}). 
We estimate errors on $\rho_L$ with the same method used on $\phi^*$ and $\bar\rho$.

We have thus far neglected the effects of photometric errors. This is
actually a reasonable simplification, as $99\%$ of our sample galaxies
have estimated errors of $< 0.1$ mag. Assuming a gaussian magnitude
error distribution with dispersion $\sigma = 0.1$ mag, we can correct
for the effects of photometric errors on the luminosity function using the method
described in EEP \S~3.5. We find that neglecting photometric errors only biases
$M^*$ and $\alpha$ by at most $\Delta M^* = -0.02$ and
$\Delta \alpha = -0.01$, much smaller than our 1$\sigma$
uncertainties (Table~\ref{tablf}). We can thus safely neglect the
effects of photometric errors in the rest of this paper.

Finally, note that for galaxies within our adopted filter redshift
limits, the magnitude selection weights $w_m$ will be slight 
{\em over}estimates, for the following reason. Recall that $w_m$ is
basically the ratio (at a given apparent magnitude $m$) of the total
number of galaxies to the number of galaxies with redshifts. However,
since we include galaxies both inside and outside the filter redshift
limits when computing this ratio, and since the redshift success rate
will be lower for galaxies outside the filter limits than
for those inside, $w_m$ will be overestimated for those galaxies
inside the limits. The amount of overestimate can be approximately
calculated once some simplifying assumptions are made. 
We adopt specific values for $M^*$ and $\alpha$ (those we find below in
\S~\ref{results}; we further assume that the luminosity function does not
change with redshift), and we take our redshift sampling rate to be
some constant within a given set of filter redshift limits.
It turns out that the effect is approximately independent
of apparent magnitude, so that the luminosity function shape is not
affected. The luminosity function normalization will, however, be too
large by roughly $20 \pm 10$\% for our adopted filter limits, where
the uncertainty represents the range of scatter observed for different
apparent magnitudes and filters.
We do not explicitly include this systematic
correction in our luminosity function normalizations, as it is
comparable to our random errors, as well as somewhat complicated to deal
with exactly (e.g., we need to model the change of luminosity function
with redshift, as well as iterate the luminosity function and
``renormalization'' calculations until convergence). 
However, we will need to keep the approximate 20\%
correction in mind when comparing our results to those of other
surveys, as we do below.

\section{Results} \label{results}

\begin{figure}
\plotone{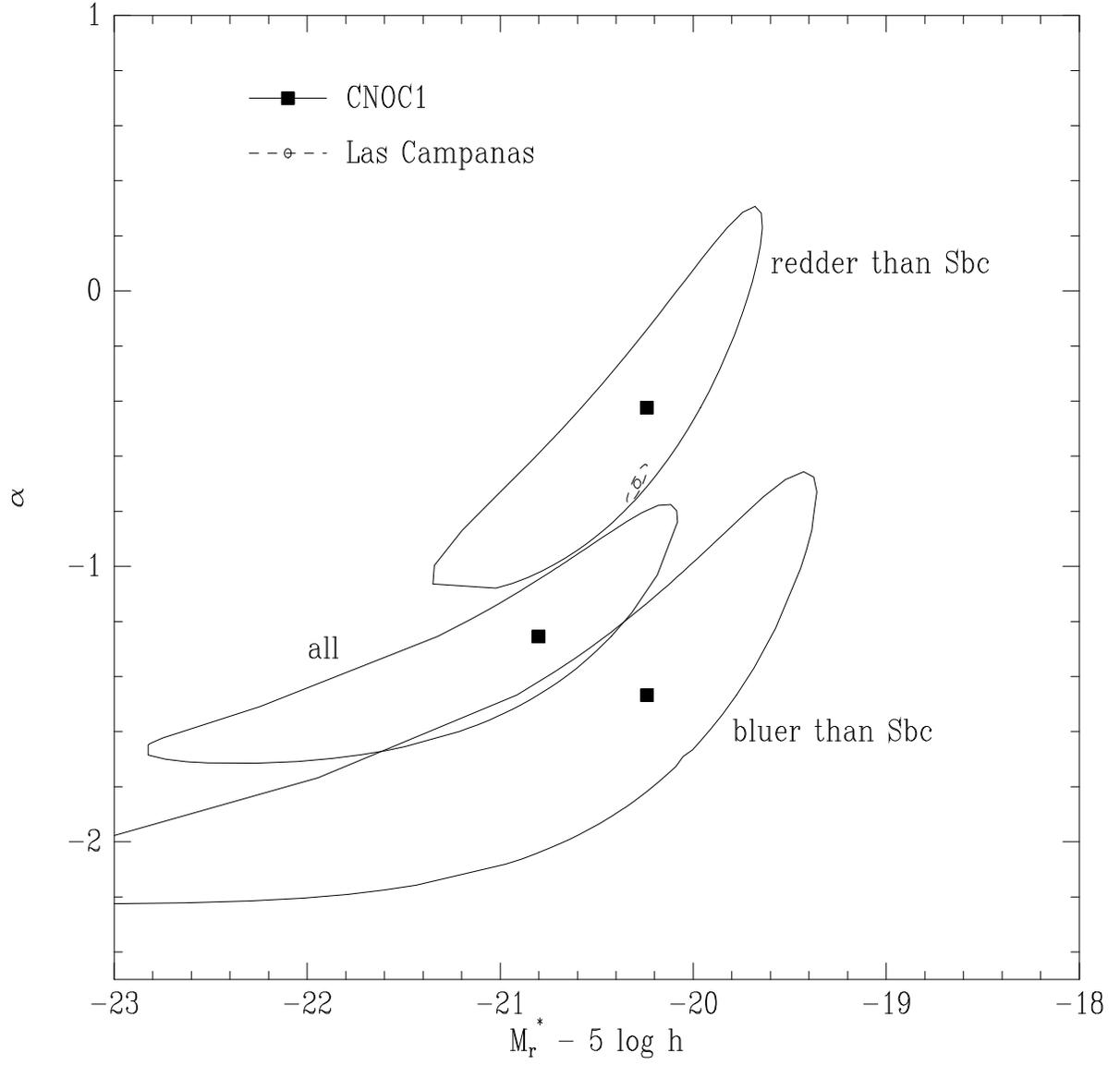}
\caption{$2\sigma$ error ellipses in $M^*$ and $\alpha$ for 
  the $r$-band luminosity functions from the CNOC1
  and Las Campanas redshift surveys.}
\label{figcontrbred}
\end{figure}

\begin{figure}
\plotone{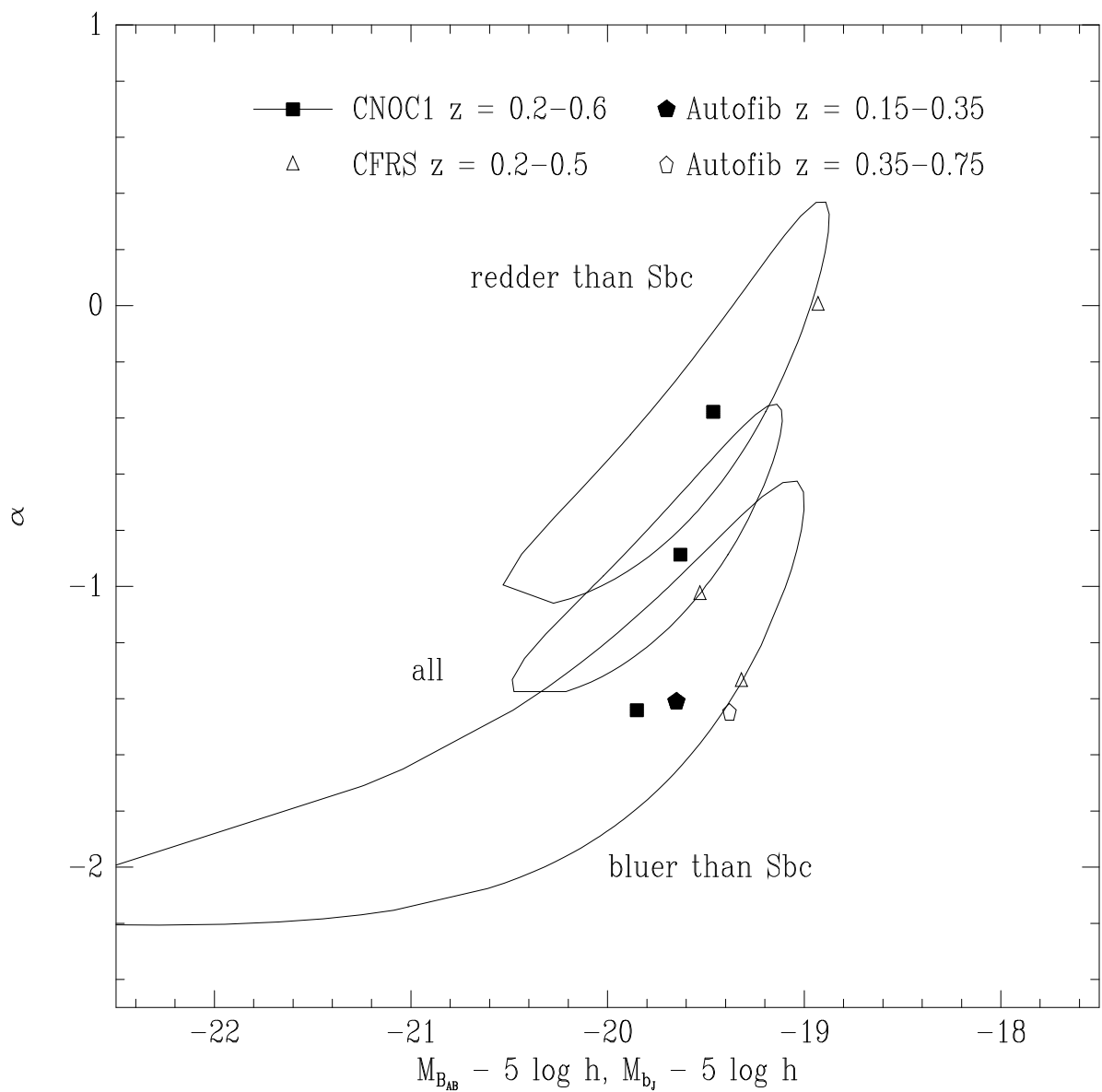}
\caption{$2\sigma$ error ellipses in $M^*$ and $\alpha$ for
  the $B$-band luminosity functions of the CNOC1 samples.
  Also shown are $M^*$ and $\alpha$ values for samples from the 
  Canada-France (CFRS) and Autofib redshift surveys. Results for red
  and blue galaxy subsamples are shown only for CNOC1 and CFRS.}
\label{figcontrbblue}
\end{figure}

\begin{figure}
\plotone{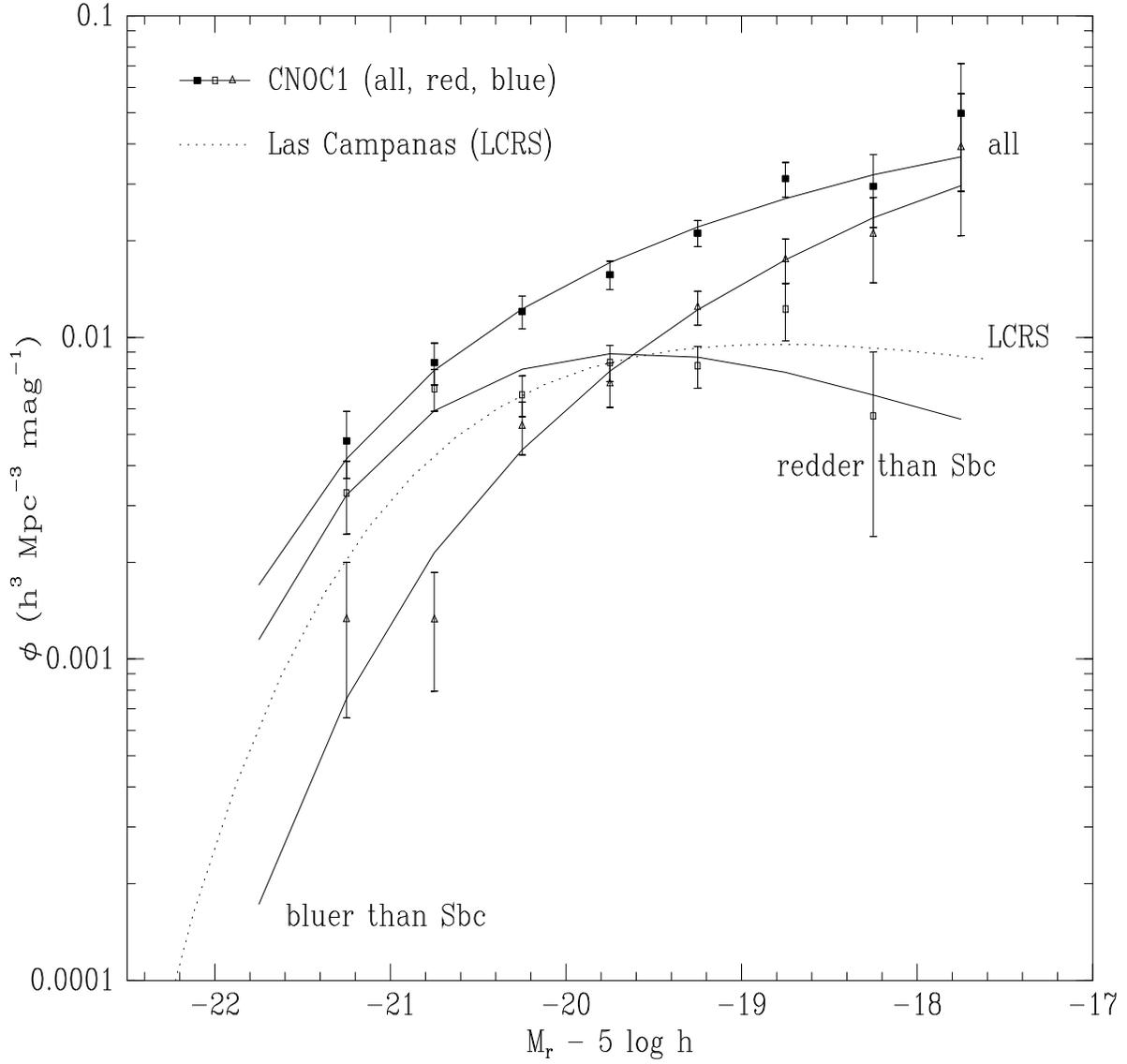}
\caption{The $r$-band luminosity function of CNOC1 samples. The lines are
  the STY fits, while the points are SWML solutions with
  $1\sigma$ errors. Also shown is the STY fit from the Las Campanas survey.}
\label{figphimrb}
\end{figure}

\begin{figure}
\plotone{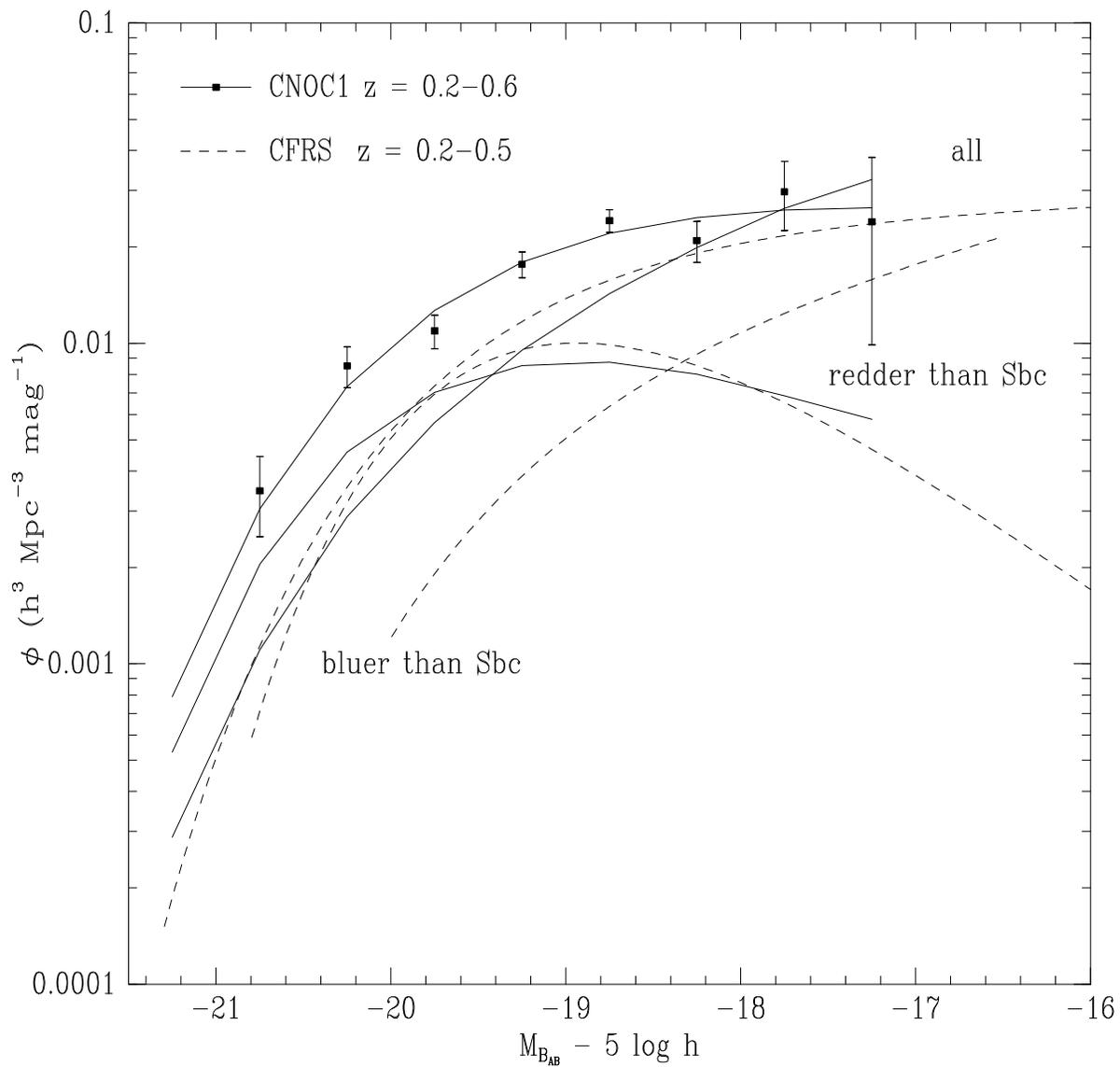}
\caption{The $B_{AB}$-band luminosity functions of CNOC1 samples are compared to
  those of corresponding CFRS samples. The points show SWML solutions
  for the CNOC1 all sample.}
\label{figphimcfrs}
\end{figure}

Our luminosity function fit results are given in Table~\ref{tablf},
for both rest-frame Gunn $r$ and rest-frame $B_{AB}$.
Figures~\ref{figcontrbred} and \ref{figcontrbblue} show 
the $2\sigma$ error ellipses in the $M^*$
and $\alpha$ parameters, and Figures~\ref{figphimrb} and
\ref{figphimcfrs} plot the actual
luminosity functions $\phi(M)$. The results are shown for the full 389
galaxy sample, as well as for two subsamples of galaxies whose
rest-frame colors are either redder (209 galaxies) or bluer (180
galaxies) than that of an Sbc galaxy (model of \cite{cww}).
Overall, the full sample luminosity functions have Schechter parameters 
$M^*_r = -20.8$ and $\alpha = -1.3$ in Gunn $r$, and
$M^*_r = -19.6$ and $\alpha = -0.9$ in $B_{AB}$.
Note the clear distinction between the red- and blue-subsample luminosity
functions shown in Figures~\ref{figphimrb} and \ref{figphimcfrs}: the
red subsample has a shallow $\alpha = -0.4$ and dominates the galaxy
population at the bright end ($M_r, M_{B_{AB}} \lesssim -19.5$), 
while the blue subsample has a much
steeper $\alpha = -1.4$ to $-1.5$, and consequently dominates at the faint end 
($M_r, M_{B_{AB}} \gtrsim -19.5$).
The error ellipses for the red and blue subsamples are
clearly separated for both Gunn $r$ and $B_{AB}$, as seen in 
Figures~\ref{figcontrbred} and \ref{figcontrbblue}, respectively.

We next compare our results against 
those from the two other largest surveys with $B$-band luminosity
functions available at comparable intermediate redshifts, specifically the 
Canada-France Redshift Survey (CFRS; \cite{lil95}) and 
the Autofib Redshift Survey (\cite{ell96}). 
We start with the CFRS.
In Figures~\ref{figcontrbblue} and ~\ref{figphimcfrs} we plot the 
CFRS $B_{AB}$-band results for a sample of 208 galaxies in 
their $z = 0.2-0.5$ bin (their ``best'' estimates). The CFRS
sample has also been divided into red and blue subsamples at the same Sbc
color cut used in CNOC1, and we show the corresponding 
results as well. 
Note first from Figure~\ref{figcontrbblue} that the CFRS 
$M^*$ and $\alpha$ values are just outside the
respective CNOC1 $2\sigma$ error ellipses. Remembering that
the corresponding CFRS error ellipses will be larger (the CFRS sample
is about half the size of CNOC1), we find that the CNOC1 and CFRS
$M^*$ and $\alpha$ values agree at better than the 2$\sigma$ level
for each of the all, red, and blue samples.
However, when we plot the actual luminosity functions in
Figure~\ref{figphimcfrs}, we find that although the {\em shapes}
(i.e., $\alpha$) of corresponding CNOC1 and CFRS luminosity functions 
agree fairly well, the CNOC1 results tend to show brighter $M^*$
values and higher normalizations, especially in the case of the blue
subsample. Recall that the CNOC1 normalization is likely an upper
bound and may have to be decreased by about 20\%, which would help
improve the match with CFRS. An additional decrease in the CNOC1 $M^*$
by 0.2 mag (cf. Figure~\ref{figcontrbblue}) would be needed to bring the
CNOC1 and CFRS full-sample luminosity functions into very close agreement.

We consider next some possible causes for the differences, though none
of them are entirely satisfactory.
Extinction, which has been neglected in our analysis, should only have
a small effect (as our photometry was in the $r$-band) and would
actually brighten our $M^*$ values. Large-scale structure may bias the
CNOC1 normalization high, as our field samples are in the neighborhood
of rich clusters. 
However, we find that our results are little changed even
when we triple the adopted cluster redshift limits (from about $\pm
3000$~km~s$^{-1}$ to $\pm 9000$~km~s$^{-1}$ from the cluster centers) 
in order to reduce the effects from potential large-scale structure 
associated with the clusters. 
Moreover, one expects the discrepancies to be
larger for red galaxies (as they are more common in dense
environments), but the observed differences are actually larger for blue
galaxies. Also, the choice of spectral energy distributions used here to
calculate $k$-corrections and to make the red/blue cuts is the same as
that in CFRS. We also verified that we could reproduce the CFRS
luminosity function results using our code on their redshift catalog data. 
A final possibility is that there remains some unaccounted systematic offset
of order 0.2 mag in the photometry zero-points, arising from the 
completely independent ways the photometry was obtained and reduced in the two
surveys. 
In the end, the
differences may simply reflect actual sampling fluctuations. It is nevertheless
encouraging that the luminosity function shapes of corresponding CNOC1
and CFRS samples are quite similar. Also, even before the 20\% correction to
the CNOC1 normalization, the luminosity densities from the two 
surveys do not differ at more than the 2$\sigma$ level (see also
below). Overall the
CNOC1 and CFRS results are roughly consistent, showing the same
luminosity function shapes, but
with normalization and $M^*$ differences, both at about the 20\% level.
We note that the CNOC1 sample provides somewhat 
better luminosity function measurements in the range $z = 0.2-0.6$
because of its larger sample size. On the other hand, the {\em full} CFRS sample
extends to $z \approx 1$ and provides a longer baseline for studying
the evolution of red and blue galaxy luminosity functions, a point we
return to below. 

\begin{figure}
\plotone{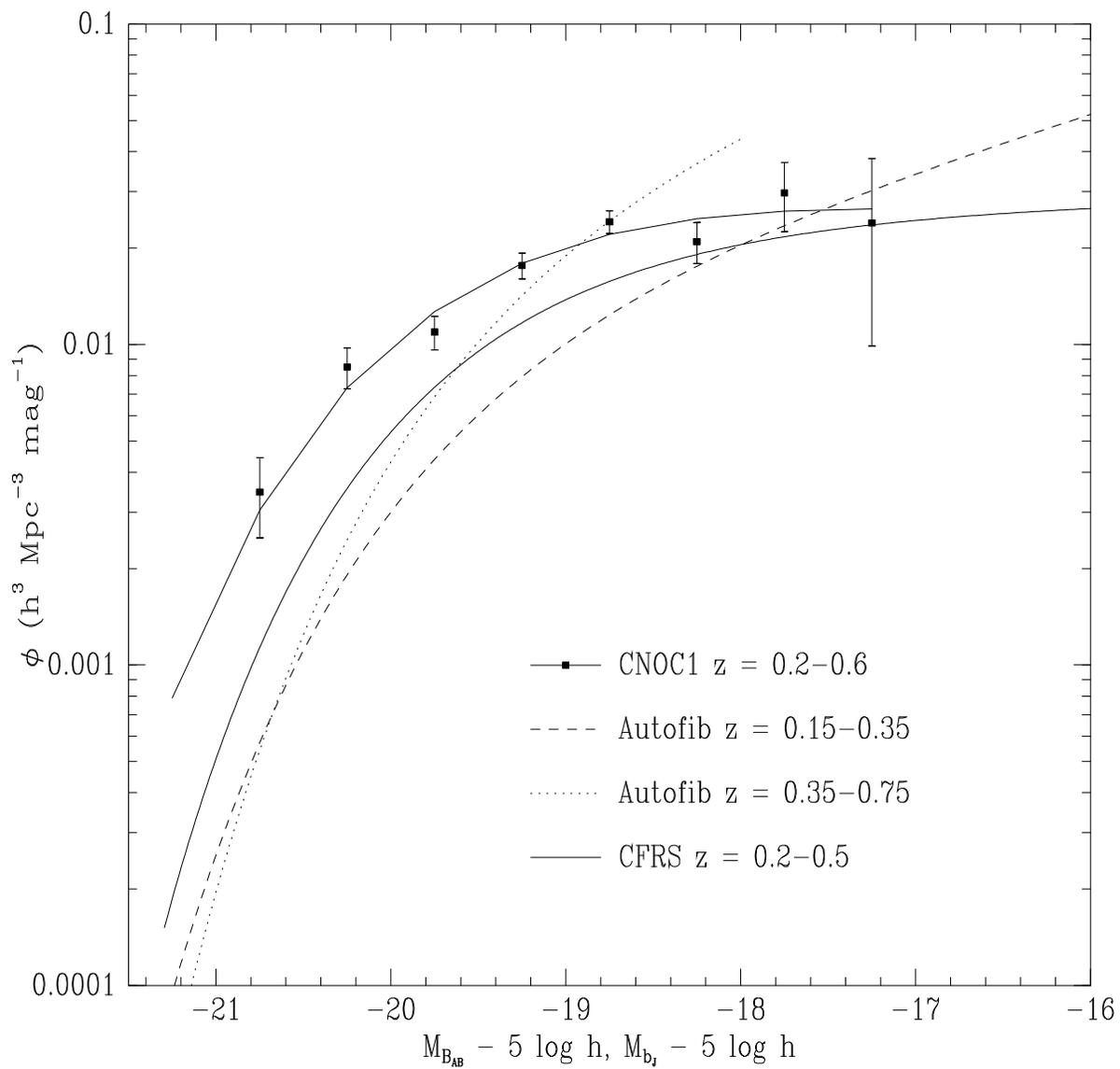}
\caption{The $B$-band luminosity function of the CNOC1 all 
  sample is compared to that of two samples from the Autofib 
  redshift survey and to that of the $z = 0.2-0.5$ all sample from 
  the CFRS.}
\label{figphimauto}
\end{figure}

Next, in Figures~\ref{figcontrbblue} and \ref{figphimauto} 
we compare our results against the 
$z = 0.15-0.35$ and $z = 0.35-0.75$ luminosity functions from the Autofib
survey. (The Autofib survey results are in the $b_J$ system, which is
close enough to $B_{AB}$ that we will not apply any zero-point corrections;
cf. \cite{fuku95}. Also, there are no results for red and blue
subsamples.) Here the agreement is worse than it was with CFRS.
Both Autofib samples show steeper faint-end slopes than those of CNOC1
and CFRS. The $M^*$ and $\alpha$ values are discrepant at the 2 to
3$\sigma$ level compared to CNOC1 (the corresponding error ellipse for
the Autofib $z = 0.15-0.35$ sample is smaller than that of the CNOC1
all sample, while that for $z = 0.35-0.75$ is comparable in size
to that of CNOC1; see Figure~11 of \cite{ell96}). Examination of
Figure~\ref{figphimauto} shows that the CNOC1, Autofib,
and CFRS luminosity functions are only broadly consistent with each
other, with Autofib samples showing the steepest slopes and lowest 
normalizations. It is curious that the Autofib results actually
resemble that of our blue subsample well (though this is {\em not} to
suggest that they are incomplete in red galaxies). 
Some of the same points raised above in discussing the
CNOC1 and CFRS differences apply here as well. Also, though the
Autofib sample is the largest among the three surveys, there are some
differences in their survey construction compared to CNOC1 and CFRS. 
Unlike CNOC1 or CFRS, the Autofib sample is actually a combination of
a number of disparate redshift surveys, carried out with different 
instruments and telescopes. Instead of CCD photometry, the Autofib
survey is based on photographic plate photometry.
Also, the original photometry is in the $b_J$ band,
rather than at longer wavelengths like $r$ (CNOC1) or $I$ (CFRS). 
Consequently the $k$-corrections needed to convert to rest-frame $B$
are largest for the Autofib samples. Ellis et al.\ (1996) and
references therein detail a careful treatment of these various issues.
While we do not claim that any of these is the cause of the luminosity
function differences, we do suggest that the details of the construction and
selection of galaxy samples from one survey to another are
complicated enough that they may systematically affect comparison of results
among different samples. The less than ideal agreement shown in
Figure~\ref{figphimauto} indicates that even larger samples, with well-defined
selection criteria, are needed to make a definitive determination of
the luminosity function at intermediate redshifts.

Now, we proceed to examine the issue of galaxy evolution within the
CNOC1 sample.
We have divided the CNOC1 samples by redshift
to look for changes in $M^*$ and $\alpha$ as a
function of $z$. We find there to be no significant changes in $M^*$
or $\alpha$ with redshift, within the respective $2\sigma$ error ellipses
of the $z$-divided samples.
This holds for the full sample, as well as for the red and blue
galaxy subsamples. However, because the error ellipses do get quite
large when the redshift cuts are applied, we are not able to
rule out (or in) changes in $M^*$ of order 0.5 magnitude over $0.2 < z
< 0.6$.

We will instead examine changes in the galaxy luminosity density
$\rho_L$ as a function of $z$, and adopt the simple procedure of
fixing $M^*$ and $\alpha$ at the values given in
Table~\ref{tablf} for all $z$. We plot the $B_{AB}$-band luminosity density
$\rho_L$ vs. $z$ in four redshift bins in
Figure~\ref{figld}, for the all, red, and blue samples.
(To facilitate comparison with other surveys, we have rescaled our
$\rho_L$ results for $M_{\rm faint} = -17 + 5 \log h$ to $M_{\rm
  faint} = +\infty$; recall equation~[\ref{eqsl}].)
The errors plotted are $1\sigma$ estimates as described in \S~\ref{methods}, with
roughly equal contributions from bootstrap resampling and density
fluctuation uncertainties. We see from the figure that $\rho_L$ for
the full sample rises at higher $z$, and that this increase is caused 
predominantly by the blue galaxy population. The red sample $\rho_L$
does not exhibit a trend with redshift, while the blue sample $\rho_L$
rises steadily with $z$. 
If we use just two bins (for simplicity), $z = 0.2-0.4$ and $z = 0.4-0.6$, 
the red galaxy $\rho_L$ in the high-$z$ bin is just 10\% more than that in the 
low-$z$ bin, while the blue galaxy $\rho_L$ nearly triples, albeit
with large errors.
These results may be directly compared to the
luminosity density vs. redshift results derived from the CFRS
(\cite{lil96}) which we also show in Figure~\ref{figld}.
We find that the {\em shapes} of the CNOC1
$\rho_L$ vs. $z$ relation are fairly similar to the CFRS
results, which were derived over a longer redshift baseline. However,
the CNOC1 luminosity densities appear systematically higher,
especially for the blue sample.
If we apply the likely 20\%
normalization reduction for CNOC1 discussed before, the results agree
better, as also shown in Figure~\ref{figld}. 
In any event, 
the CNOC1 trends are at least qualitatively consistent with those seen in the CFRS:
strong evolution of the luminosity density of blue galaxies, but little 
evolution of that of red galaxies. 
Moreover, Schade et al.\ (1996) have found that the
surface brightness of disk galaxies in the CNOC1 survey increases with
$z$ in a way consistent with the shape of the CFRS $\rho_L$-$z$
relation; our current results thus also corroborate the earlier CNOC1 findings.
That the differences between CNOC1 and CFRS lie
in the absolute normalizations rather than in the shapes suggests
that {\em within} each survey, we are basically measuring the same galaxy
luminosity density trends, but that there is some unaccounted systematic 
normalization or scaling difference, particularly for blue galaxies, 
which causes the disagreement when we compare {\em across} the two surveys. 

We have also checked that our red/blue luminosity function and luminosity
density evolution differences are robust to potential errors in
assigning galaxies to the red and blue samples (e.g., a potential
wavelength-dependent error in our model Sbc galaxy spectrum can lead
to a redshift-dependent type assignment error). We do this check by
more finely dividing our sample into three roughly equal parts by color.
We find that the steep faint end of the blue luminosity
function and the increase of the blue luminosity
density at higher redshift both result primarily from galaxies in the
bluest third of our sample, rather than from blue galaxies near the
Sbc dividing line. In the latter case our results would have been more
sensitive to potential type-assignment errors, but as the former case is
the actual situation, our conclusions regarding the red/blue differences 
should not be affected.

Finally, we compare our $r$-band results against those from the {\em local} sample of 18,678 
Las Campanas Redshift Survey (LCRS) galaxies, the largest galaxy
sample for which the luminosity function has been computed
(\cite{lin96lum}). The LCRS galaxies have an average redshift $z = 0.1$, and
are also observed in the red\footnote{Specifically an isophotal,
  ``hybrid'' Kron-Cousins $R$-band (details in \cite{tuck94} and
  \cite{lin96lum}), whose zero-point is within $\approx 0.1-0.2$ mag of
  the Gunn-$r$ system used for CNOC1.}. From Figures~\ref{figcontrbred}
and \ref{figphimrb}, we find that the local LCRS luminosity function
is clearly different from the intermediate redshift CNOC1 results.
The CNOC1 results show a brighter $M^*$, a steeper faint-end slope
$\alpha$, as well as a higher normalization compared to the LCRS.
We note that if the differences result purely from galaxy evolution, that
evolution needs to to be luminosity dependent as well, as the
differences become larger at the faint end of the luminosity function.
In terms of luminosity
densities, the LCRS has $\rho_L = (1.9 \pm 0.2) \times 10^8 \ L_{\sun} \
h$~Mpc$^{-3}$, about half that of the CNOC1
sample. The increase in luminosity density at intermediate
redshifts relative to the local value appears to be a robust conclusion, as
the CNOC1 results are corroborated by independent samples like CFRS
and Autofib (despite the differences among the intermediate-redshift samples).
However, a careful consideration and account of systematic effects
will be especially important for a definitive assessment of the causes
of the differences.
For example, different surface brightness limits of local vs. deeper
surveys (\cite{fer95}), or different photometry methods 
(different aperture sizes, limiting isophotes, etc.) can cause 
differences in the luminosity function that have nothing to do with 
actual galaxy evolution. 
Though beyond the scope of this paper, such
a detailed modelling of both evolutionary and observational effects 
is planned for the ongoing CNOC2 redshift survey, a successor of 
CNOC1, which
should eventually provide a much larger sample of several thousand
intermediate redshift field galaxies for study. 

\section{Conclusions} \label{conclusions}

We have computed the rest-frame Gunn $r$ and $B_{AB}$ luminosity
functions for a sample of 389 
field galaxies from the CNOC1 redshift survey, over the redshift
range $z = 0.2-0.6$. We find Schechter parameters 
$M^*_r - 5 \log h = -20.8 \pm 0.4$ and 
$\alpha = -1.3 \pm 0.2$ in Gunn $r$, and
$M^*_{B_{AB}} - 5 \log h = -19.6 \pm 0.3$ and 
$\alpha = -0.9 \pm 0.2$ in $B_{AB}$.
Samples of red and blue galaxies, cut at the rest-frame $g-r$ color of
an Sbc galaxy, show distinctly
different luminosity functions. Red galaxies have a shallow slope
$\alpha \approx -0.4$ and dominate the bright end of the luminosity
function, while blue galaxies have a steep $\alpha \approx -1.4$ and prevail
at the faint end. Table~\ref{tablf} summarizes our results.

\begin{figure}
\plotone{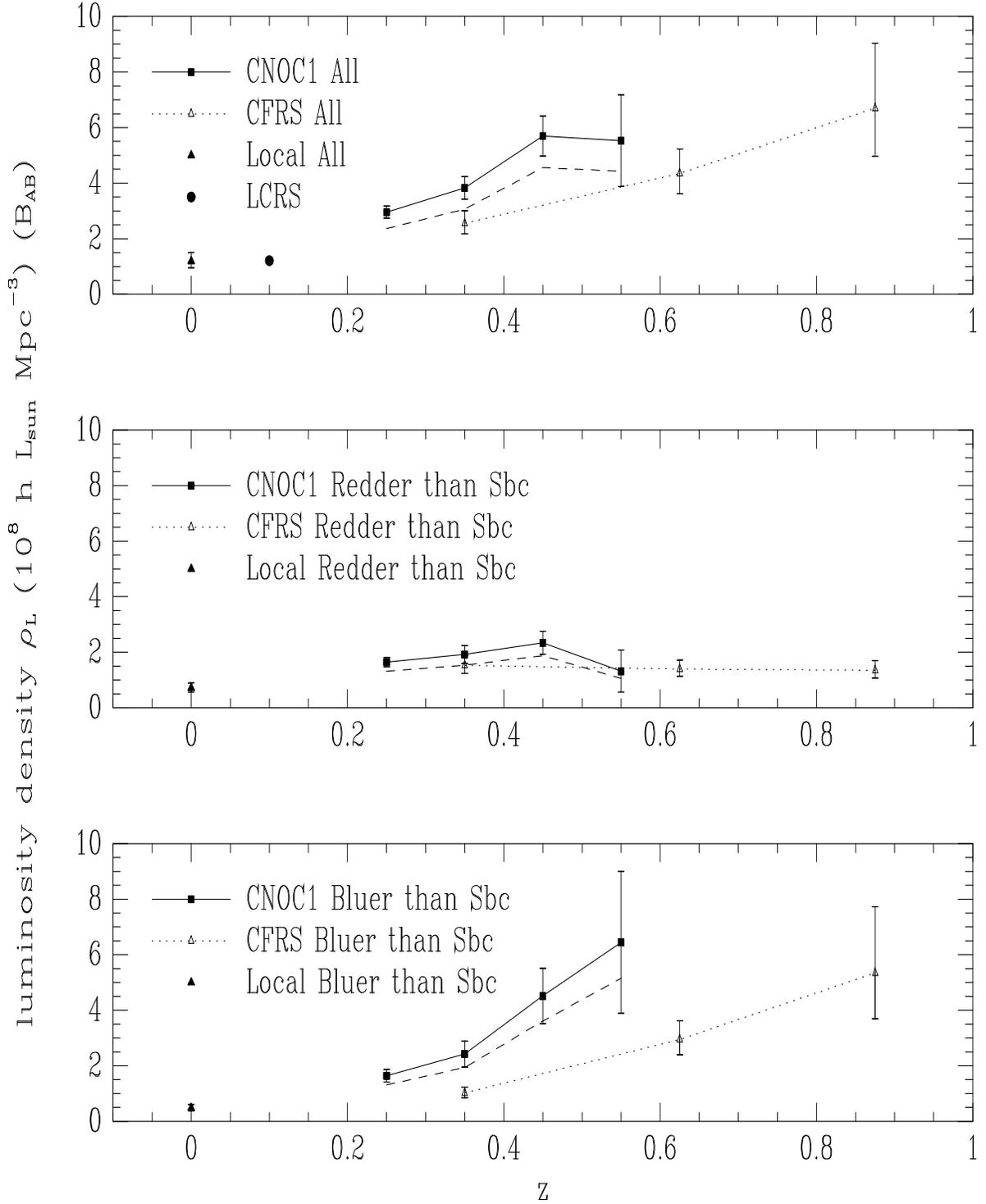}
\caption{The $B_{AB}$-band luminosity density $\rho_L$ for CNOC1 samples, compared
  to that from analogous CFRS samples. The dashed lines show the CNOC1
  results after including a 20\% reduction in normalization as
  discussed in the text. Note that the CNOC1 and CFRS curves differ
  primarily in their absolute {\em normalization}, particularly for
  the blue subsample. The {\em shapes} of the
  trends are very similar, once we rescale to match the 
  normalizations. Also shown are results for local surveys as
  estimated by Lilly et al. (1996), and in the top panel, the result from the
  LCRS, converted to the $B_{AB}$-band using $B_{AB} - R_{LCRS} = 1$.}
\label{figld}
\end{figure}

Comparisons of the CNOC1 results to independently-determined 
intermediate-redshift 
luminosity functions show broad agreement with results from the 
CFRS and Autofib redshift surveys. However, there are questions about
particular differences in the luminosity functions which will require larger 
samples to resolve. 
Calculation of the $B_{AB}$-band luminosity density shows that the CNOC1 
red galaxy luminosity density is about the same from $z = 0.2-0.6$,
but that the blue luminosity density over $z = 0.4-0.6$ is nearly
three times that for $z = 0.2-0.4$.
These trends are consistent with those derived from the CFRS 
(see Figure~\ref{figld}), except for a normalization difference resulting
primarily from blue galaxies,
and are also consistent with the
redshift evolution of disk galaxy surface brightness observed
in CNOC1. Comparison to the local luminosity function from
the Las Campanas redshift survey implies that the luminosity density at
$z \approx 0.1$ is only about half that seen at $z \approx 0.4$. Also,
luminosity-dependent evolution, increasing at the faint end, 
would seem to be required to match the
CNOC1 and LCRS luminosity functions, if galaxy evolution is the sole
cause of the differences seen.

However, an underlying caveat throughout our comparisons of CNOC1
results with those of other surveys is that the particular details of
the construction of different surveys may have an important effect on
how we interpret the results. Along these
lines, we have pointed out some potential systematic causes of the differences
among the various survey luminosity functions we considered, though we
have not attempted a detailed investigation. It is probably fair to
say that control of systematic effects within a single survey is
simpler than across different surveys, and perhaps that is why the
red/blue luminosity density trends appear to have the same shapes but 
different normalizations in CNOC1 and CFRS.
In any event, the CNOC1 field sample is still relatively small, and a
larger sample will
definitely help confirm or reject some of the trends that we have seen
in CNOC1, as well as help resolve the differences seen vs. other surveys. 
In particular, the CNOC2 field redshift survey should
provide us with an order of magnitude larger sample of several
thousand objects to examine the luminosity and other properties of
galaxies at intermediate redshifts. We thus anticipate a similar but more detailed
analysis for the CNOC2 data in the near future, and a thorough
investigation of the evolution of galaxies from intermediate to low redshifts.

\acknowledgments
We thank the many CNOC team members who participated in the
observation and reduction of data for the survey. 
We thank Simon Lilly and David Schade for helpful discussions and for
providing spectral energy distributions and related information. We
also thank the referee David Koo for helpful suggestions.
Financial support from NSERC and NRC of Canada is also gratefully acknowledged.

\clearpage

\begin{deluxetable}{crcc}
\tablewidth{0pt}
\tablecaption{Sample Parameters}
\tablehead{
   \colhead{Filter\tablenotemark{a}}  
 & \colhead{$N$\tablenotemark{b}}  
 & \colhead{Redshift Limits} 
 & \colhead{Apparent Magnitude Limits \tablenotemark{c}}
 }
\startdata
 Z2 &  99 & $0.20-0.27$ & $18.0-20.5$ \nl
 Z3 &  45 & $0.27-0.38$ & $19.0-21.5$ \nl
 Z4 & 217 & $0.21\tablenotemark{d}-0.43$ & $19.0-21.7$ \nl
 Z5 &  27 & $0.45-0.60$ & $20.0-22.0$ \nl
\tablenotetext{a}{Names of band limiting filters used for
  spectroscopy; see text and YEC for further details.}
\tablenotetext{b}{The samples are further restricted to the
  Gunn $r$ absolute magnitude range $-22.0 < M_r - 5 \log h < -17.5$. The
  numbers remain the same for the alternative restriction
  $-21.5 < M_{B_{AB}} < -17.0$, except that there is
  one additional galaxy in the Z4 sample.}
\tablenotetext{c}{Gunn $r$ magnitudes.}
\tablenotetext{d}{The value given in YEC was 0.27 and was based only on
  detectability of [OII] $\lambda 3727$. The present value is extended 
  down to 0.21 based on the added detectability of 
  [OIII] $\lambda\lambda 5007,4959$ at those low redshifts where 
  [OII] $\lambda 3727$ would disappear off the blue end of a
  spectrum. Experience has shown that in the CNOC1 sample the [OII]
  and [OIII] features are always detected simultaneously whenever they
  are all potentially visible.}
\label{tabsamps}
\enddata
\end{deluxetable}

\clearpage

\begin{deluxetable}{crcccl}
\tablewidth{0pt}
\tablecaption{Luminosity Function Parameters \tablenotemark{a}} 
\tablehead{
   \colhead{Sample} 
 & \colhead{$N$} 
 & \colhead{$M^* - 5 \log h$ \tablenotemark{b}}
 & \colhead{$\alpha$ \tablenotemark{b}} 
 & \colhead{$\phi^*$ \tablenotemark{c,e}} 
 & \colhead{$\rho_L$ \tablenotemark{d,e}}
 }
\startdata
Gunn $r$ & & & & & \nl
 & & & & & \nl
All   & 388 & $-20.80 \pm 0.40$ & $-1.25 \pm 0.19$ & $0.021 \pm 0.008$
      & $4.3 \pm 0.5$ \nl
Red   & 209 & $-20.24 \pm 0.30$ & $-0.42 \pm 0.28$ & $0.024 \pm 0.005$
      & $2.3 \pm 0.3$ \nl
Blue  & 179 & $-20.24 \pm 0.52$ & $-1.47 \pm 0.32$ & $0.013 \pm 0.009$ 
      & $1.8 \pm 0.4$ \nl
 & & & & & \nl
 & & & & & \nl
$B_{AB}$ & & & & & \nl
 & & & & & \nl
All   & 389 & $-19.63 \pm 0.25$ & $-0.89 \pm 0.21$ & $0.042 \pm 0.010$
      & $3.6 \pm 0.4$ \nl
Red   & 209 & $-19.46 \pm 0.30$ & $-0.38 \pm 0.29$ & $0.024 \pm 0.005$
      & $1.8 \pm 0.3$ \nl
Blue  & 180 & $-19.85 \pm 0.50$ & $-1.44 \pm 0.32$ & $0.014 \pm 0.009$ 
      & $2.0 \pm 0.4$ \nl
\tablenotetext{a}{Absolute magnitude restrictions 
  $-22.0 < M_r - 5 \log h < -17.5$ for Gunn $r$, and
  $-21.5 < M_{B_{AB}} - 5 \log h < -17.0$ for $B_{AB}$, are applied in the
  definitions of the samples and the calculations of the luminosity
  function and associated parameters, except that $\rho_L$ is computed
  for $-\infty < M - 5 \log h < -17$.} 
\tablenotetext{b}{Errors are $1\sigma$ {\em one}-parameter errors
  determined from the STY fits. See Figures~1 and 2 for the
  full {\em two}-parameter $M^*$-$\alpha$ error ellipses.}
\tablenotetext{c}{Units are $h^3$~Mpc$^{-3}$~mag$^{-1}$. Errors are
  $1\sigma$ and are estimated as described in the text.}
\tablenotetext{d}{Units are $10^8 \ L_{\sun} \ h$~Mpc$^{-3}$. We take
  Gunn $r_{\sun} = 4.84$ and $B_{AB \sun} = 5.34$, as inferred from solar
  photometric data from Allen 1973, \S~75, and photometric
  transformations from Kent 1985 and Fukugita et al.\ 1995. 
  Errors are $1\sigma$.}
\tablenotetext{e}{Note that $\phi^*$ and $\rho_L$ should likely be
  reduced by about 20\%, due to the systematic effect discussed at
  the end of \S~3.}
\label{tablf}
\enddata
\end{deluxetable}

\clearpage

\clearpage

\clearpage

\clearpage

\clearpage

\clearpage

\end{document}